\documentclass[aps,prl,groupedaddress,showpacs,twocolumn]{revtex4}

\usepackage{graphicx}
\usepackage{amsmath}
\usepackage{amssymb}
\usepackage{bm}

\begin{document}
\title{Phase diagram of spin 1 antiferromagnetic Bose-Einstein condensates}
\author{David Jacob, Lingxuan Shao, Vincent Corre, Tilman Zibold, Luigi De Sarlo, Emmanuel Mimoun, Jean Dalibard and Fabrice Gerbier\email{fabrice.gerbier@lkb.ens.fr}}

\affiliation{Laboratoire Kastler Brossel, CNRS, ENS, UPMC, 24 rue Lhomond, 75005 Paris}
\affiliation{}
\date{\today}
\begin{abstract}
We study experimentally the equilibrium phase diagram of a spin 1 Bose-Einstein condensate with antiferromagnetic interactions, in a regime where spin and spatial degrees of freedom are decoupled. For a given total magnetization $m_z$, we observe for low magnetic fields an ``antiferromagnetic"  phase where atoms condense in the $m=\pm 1$ Zeeman states, and occupation of the $m=0$ state is suppressed. Conversely, for large enough magnetic fields,  a phase transition to a ``broken axisymmetry" phase takes place: The $m=0$ component becomes populated and rises sharply above a critical field $B_c(m_z)$. This behavior results from the competition between antiferromagnetic spin-dependent interactions (dominant at low fields) and the quadratic Zeeman energy (dominant at large fields). We compare the measured $B_c$ as well as the global shape of the phase diagram with mean-field theory, and find good quantitative agreement.
\end{abstract}
 \maketitle
%
%
One of the most active topics in the field of ultra cold quantum gases is the study of interacting many-body systems with spin \cite{ho1998a,ohmi1998a,stamperkurn2012a}. Atoms with arbitrary Zeeman structure can be trapped using far-detuned optical traps. Quantum gases of fermions with spin larger than $1/2$ \cite{taie2010a,krauser2012a} and bosons with spin $1$ \cite{stenger1998a,chang2004a,schmaljohann2004a,chang2005a,kronjaeger2005a,black2007a,liu2009a,saddler2006a,bookjans2011a,guzman2011a}, $2$ \cite{schmaljohann2004a,kuwamoto2004a}, or $3$ \cite{pasquiou2012a} have been demonstrated experimentally. This opens a whole class of new experiments with spinful many-body systems, such as observation of squeezing among the different spin components, coherent spin mixing dynamics analogous to an internal Josephson effect \cite{chang2004a,schmaljohann2004a,kuwamoto2004a,chang2005a,kronjaeger2005a,black2007a,liu2009a}, or the study of sudden quenches across magnetic phase transitions \cite{saddler2006a,bookjans2011a}. 

The simplest example is the spin-1 Bose gas, where the spin-dependent interactions can favor either ferromagnetic (the case of atomic $^{87}$Rb \cite{chang2004a}) or antiferromagnetic (the case of atomic $^{23}$Na \cite{stenger1998a}) behavior, leading to different equilibrium phases. An additional but essential feature in experiments with gases of alkali atoms is the conservation of the longitudinal magnetization $m_z=n_{+1}-n_{-1}$; Here $n_{m}$ denotes the relative populations of the Zeeman state labeled by the magnetic quantum number $m=0,\pm 1$. This follows from the spin rotational symmetry of short-ranged two-body interactions : The only possible spin-changing two-body process is 
\begin{equation} \label{Eq:spinchanging}
m=0 ~+ m=0 \rightarrow m=+1 ~+ ~m=-1,
\end{equation}
where two $m=0$ atoms collide to yield one atom in each state $m=\pm 1$ (or vice-versa), leaving $m_z$ unchanged. For antiferromagnetic interactions, the spin energy of the right hand side in Eq.~(\ref{Eq:spinchanging}) is lower than that of the left hand side. In most physical systems, the magnetization would relax by coupling to an external environment. In contrast, quantum gases are almost perfectly isolated and the conservation of magnetization plays a major role \footnote{This statement holds when magnetic dipole-dipole interactions are negligible, which is the case for alkali atoms. For some atomic species with large magnetic moments, such as Chromium \cite{pasquiou2012a}, dipolar relaxation can be dominant.}. 

In spite of intense theoretical activity \cite{stamperkurn2012a}, the equilibrium properties of spinor gases remain relatively unexplored experimentally. Most experimental work so far have focused on dynamical properties. For ferromagnetic Rubidium condensates, a recent experimental study concluded that the time needed to reach an equilibrium state, typically several seconds or tens of seconds, could easily exceed the condensate lifetime \cite{guzman2011a}. For antiferromagnetic $^{23}$Na, the stationary regime after damping of spin-mixing oscillations has been studied for relatively high magnetization ($m_z\lesssim0.5$)\cite{liu2009a}. Here also, long equilibration times on the order of $10~$s were observed. Both experiments worked with condensates with large atom numbers, well in the Thomas-Fermi regime, where spin domains are expected and observed in transient regimes.

In this Letter, we present an experimental study of the phase diagram of spin 1 Sodium Bose-Einstein condensates with antiferromagnetic interactions. We work with small atomic samples containing a few thousands atoms held in a tightly focused optical trap; In this regime, spin domains are energetically costly, and spatial and spin degrees of freedom are largely decoupled. We prepare the sample well above the condensation temperature with a well-defined longitudinal magnetization and no spin coherence. At the end of the cooling stage, equilibration times of 3~s are used to ensure that thermal equilibrium is reached. We find, in agreement with theoretical predictions, a phase transition from an ``antiferromagnetic" phase where only the $m=\pm1$ Zeeman components are populated to a mixed ``broken axisymmetry" phase where all three Zeeman states can coexist.  We determine the phase boundary and the shape of the phase diagram versus applied magnetic field and magnetization by measuring the population of the $m=0$ state. Our measurements can be explained quantitatively by mean-field theory in the single-mode regime, where the atoms condense in the same spatial wave function irrespective of their internal state.

\begin{figure}[h!!!!!!]
\includegraphics[width=8cm]{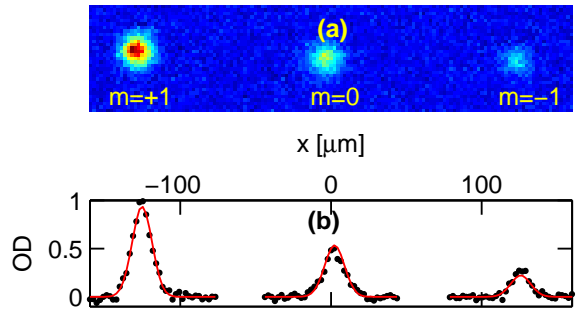}
\caption{(Color online)  {\bf a: } Absorption image of a spin 1 BEC after expansion in a magnetic gradient.  {\bf b: } Horizontal cuts through the images in {(\bf a}). The same function (shown by straight lines), only recentered and reweighted, is used to fit the density profile of each Zeeman state. }
\label{Fig:SMA}
\end{figure}

We work with Sodium atoms cooled deeply in the quantum degenerate regime using an all-optical cooling sequence \cite{mimoun2010a,jacob2011a}. In order to prepare the sample with a well-defined longitudinal magnetization and no spin coherences, we start from a cold cloud in a crossed optical dipole trap loaded from a magneto-optical trap \cite{jacob2011a}, with a magnetization $m_{z} \approx 0.6$ resulting from the laser cooling process. To obtain higher degrees of spin polarization, we perform evaporative cooling in the presence of a vertical magnetic field gradient for about 1~s. Each Zeeman state sees a slightly different potential depth. Because of the combined action of gravity and of the magnetic gradient, evaporative cooling in this configuration favors the Zeeman state with the higher trap depth \cite{couvert2009a}. This results in partially or almost fully polarized samples with magnetization up to $m_z\approx0.85$. To obtain lesser degrees of polarization than the initial value $m_z\approx0.6$, we remove the gradient and apply instead an additional oscillating field resonant at the Larmor frequency. The two procedures together allow to prepare well-defined magnetizations ranging from $0$ to $\approx 0.85$ with good reproducibility and keeping the same evaporative cooling ramp in all cases. After this ``spin distillation'', we transfer the cloud in the final crossed dipole trap and resume evaporative cooling (see the Supplemental Informations for more details about the procedure). 

After the evaporation ramp, we obtain quasi-pure spin 1 Bose-Einstein condensates  (BEC) containing $N\approx 5000$ atoms in a trap with average frequency $\overline{\omega}\approx 2\pi \times 0.7~$kHz. To ensure that the cloud has reached a steady state, we allow for an additional hold time of $3~$s after the evaporation ramp. We have investigated  the dynamics of the spin populations as this hold time is varied for several values of magnetization and applied magnetic field. We found that the populations relaxed to steady-state values with a characteristic ($1/e$) time smaller than $1~$s, much less than the finite lifetime of our sample, around $10~s$. 

The populations of the Zeeman states $m=0,\pm1$ are analyzed after expansion in a magnetic field gradient producing a Stern-Gerlach force that accelerates atoms in $m=\pm 1$ in opposite directions. After a given expansion time (typically $t\approx 3.5~$ms), we take an absorption picture of the clouds (see Fig.~\ref{Fig:SMA}a), and count the normalized populations $n_{m}$ of the Zeeman state $m=0,\pm1$. Note that the condensate is in a regime intermediate between the ideal gas and the Thomas-Fermi limits  (with  a chemical potential $\mu\approx 4\hbar\overline{\omega}$). 

For a Bose-Einstein condensate held in a tight trap as in our experiment, the energetic cost of spin domains is large (comparable to $\hbar\overline{\omega}$ per atom, much larger than the spin-dependent interaction energy). In this limit, it is reasonable to make the single mode approximation (SMA) for the condensate wavefunction \cite{law1998a,yi2002a}, which amounts to consider that all atoms share the same spatial wavefunction independently of their internal state; The condensate spin remains as degree of freedom. To support this approximation, we note that absorption images as in Fig.~\ref{Fig:SMA}a do not reveal any spatial structures or spin domains. Furthermore, we compare in Fig.~\ref{Fig:SMA}b the observed distributions with a common mode distribution. This common mode function is extracted from a Gaussian fit to the more populated cloud ($m=+1$ in this example), and then recentered and reweighed to match the populations of the other Zeeman states. We find very good agreement between the three spatial distributions in the whole range of parameters explored, and conclude that the SMA is indeed a good approximation in our case.

Because the longitudinal magnetization $m_z=n_{+1}-n_{-1}$ is conserved, the relevant magnetic energy in an applied magnetic field is the second-order (quadratic) Zeeman shift  of magnitude $q=q_B B^2$, with $B$ the applied magnetic field and $q_B \approx 277~$Hz/G$^2$; The larger (first-order) linear Zeeman shift has no influence (it can be absorbed in the Lagrange multiplier associated  to the fixed magnetization). As other spin-changing mechanisms than collisions are possible, this conservation law is only approximate. For example, it no longer holds when spin-flips are induced on purpose by applying oscillating fields as described above, or for systems with magnetic dipole-dipole interactions \cite{pasquiou2012a}. In the absence of such applied fields, we find no evidence for violation of this conservation law within our experimental limit of a few percents.


\begin{figure}[ht!!!!!!!!!!!!!!]
\begin{tabular}{rcl}
\includegraphics{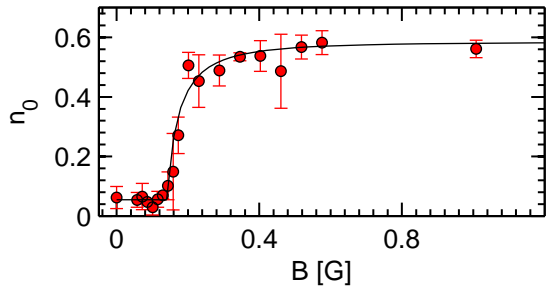} 
\end{tabular}
\caption{(Color online) Sample data showing the population $n_0$ of the $m=0$ Zeeman state versus applied magnetic field $B$, for a magnetization $m_z \approx 0.4$. The solid line is a fit to the data using Eq.~(\ref{Eq:fitfunction}). Vertical error bars show statistical uncertainties on the measured values (one standard deviation).}
\label{Fig:dataMz05}
\end{figure}

\begin{figure*}[ht!!!!!!!!!!!!!!]
\begin{tabular}{cc}
 \includegraphics{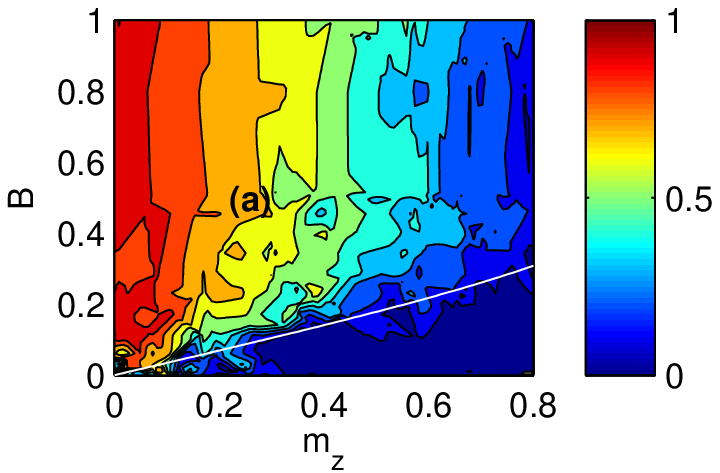}& \includegraphics{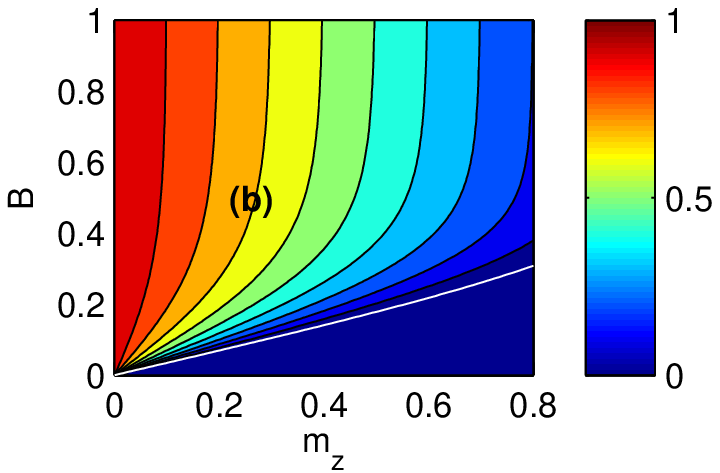}
\end{tabular}
\caption{(Color online) {\bf a: } Experimental phase diagram showing the population $n_{0}$ of the $m=0$ Zeeman state versus magnetization $m_{z}$ and applied magnetic field $B$. The plot shows a contour interpolation through all data points, with magnetization ranging from $0$ to $0.8$. The white line is the predicted critical field $B_c$ separating the two phases, deduced from Eq.~(\ref{Eq:qc}) by $q_c=q_B B_c^2$. {\bf b: } Theoretical prediction for $n_{0}$ at $T=0~$K.}
\label{Fig:PhaseDiagram}
\end{figure*}


We show in Fig.~\ref{Fig:dataMz05} the measured values of $n_{0}$ for a range of applied magnetic fields $B$ and $m_{z}\approx0.4$. The population in $m=0$ is small at low applied fields and rises sharply above a critical value $B_{c}$ before settling at an asymptotic value. We have repeated these measurements for a wide range of $B$ and $m_{z}$, and generically observed this behavior. We show the results in a reconstructed contour plot in Fig.~\ref{Fig:PhaseDiagram}a. The phase diagram shows unambiguously the presence of two different phases which differ in their spin composition, or more precisely are characterized by the absence or presence of condensed atoms in $m=0$. 


We now explain the observed behavior of $n_0$ in terms of the competition between the spin-dependent interactions and the applied magnetic field (entering quadratically through the second order Zeeman effect). The mean-field energy functional in the single-mode approximation is given by \cite{stamperkurn2012a}
\begin{eqnarray}
\label{Es}
\frac{E_{s}}{N}= \frac{U_{s}}{2} \left\vert {\bm S}  \right\vert^2 - q n_{0}.
\end{eqnarray}
Here, ${\bm S} = \langle {\bm \zeta} \vert \hat{\bm S} \vert {\bm \zeta} \rangle$ is the expectation value of the spin operator $\hat{\bm S} $ taken in the normalized spinor ${\bm \zeta}$ describing the condensate spin wavefunction, and $U_s$ denotes the spin-spin interaction energy (see Supplementary Material). For antiferromagnetic interactions ($U_{s}>0$), no applied field ($q=0$) and zero magnetization,  the spin 1 BEC realizes a polar, or ``spin-nematic'', phase according to mean-field theory \cite{ho1998a,ohmi1998a}. The spin wave function ${\bm \zeta}$ belongs to the family of eigenstates of $\hat {\bm S}\cdot {\bm n}$ with zero eigenvalue (and zero average spin), with ${\bm n}$ a headless vector called ``nematic director" in analogy with the analogous order parameter characterizing nematic liquid crystals. When $q=0$, any direction ${\bm n}$ is a possible solution, while any positive $q$ favors occupation of the $m=0$ state (along $z$) and pins the nematic director in the $z$ direction. 

When $m_z$ is non zero, there is a competition between the spin-dependent interactions and the quadratic Zeeman energy. The constraint of a fixed magnetization is essential to understand the spin structure of the condensate \cite{zhang2003a}. The BEC spin wavefunction can be parameterized generically as \cite{ho1998a,ohmi1998a,zhang2003a}
\begin{eqnarray}
{\bm \zeta}
= \left(
\begin{array}{c}
\sqrt{ \frac{1}{2}\left(1-n_0+ m_z\right)}~e^{i\phi_{+1}} \\
\sqrt{n_0}~e^{i\phi_{0}} \\
\sqrt{\frac{1}{2}\left( 1-n_0- m_z\right)}~e^{i\phi_{-1}} \end{array}\right).
\end{eqnarray}
We introduced the phases $\phi_m$ of the components of ${\bm \zeta}$ in the standard basis. The effect of antiferromagnetic spin-dependent interactions ($U_{s}>0$)  is two-fold: First, they lock the relative phase $\phi_{+1}+\phi_{-1}-2\phi_{0}$ to $\pi$ in the minimal energy state. Second, they favor the coexistence of the $m=\pm1$ component and disfavor mixing them with the $m=0$ component \cite{stenger1998a}. As the quadratic Zeeman energy favor the latter, the competition between the two results in two distinct phases as observed experimentally.

The equilibrium population $n_{0}$ is found by minimizing the mean-field energy functional \cite{zhang2003a}. For low $q$ and non-zero magnetization $m_z$, spin-dependent interactions are dominant, and result in a two-component condensate where the Zeeman states $m=\pm 1$ are populated ($n_{0}=0$). Following \cite{ueda2009a}, we will call this phase ``antiferromagnetic" (AF). When $m_{z} \rightarrow 0$, this gives an ``easy-plane" polar phase where the nematic director is confined to the $x-y$ plane. Above a critical value $q_c$ given by
\begin{eqnarray}
\label{Eq:qc}
q_{c}=U_s \left(1-\sqrt{1-m_{z}^2}\right),
\end{eqnarray} 
$n_{0}$ increases continuously from zero, indicating a second-order quantum phase transition. Again following \cite{ueda2009a}, we call this phase``broken axisymetry" (BA). For large $q$, the energy is minimized by increasing $n_0$ as much as possible given the constraint of a given $m_z$: The spin populations therefore tend to $n_{+1}=m_{z}$, $n_{0}=1-m_{z}$, $n_{-1}=0$ for $m_z>0$. When $m_{z} \rightarrow 0$, one recovers the easy-axis polar phase with all atoms in the $m=0$ state along $z$. More generally, the BA state with $n_{0}\neq 0$ has non-zero longitudinal and transverse magnetization (both vanish when $m_{z}$ goes to zero), and a nematic director orthogonal to the direction of the magnetization vector \cite{zhou2004a}. 

\begin{figure}[h!!!!!!]
\includegraphics[width=8cm]{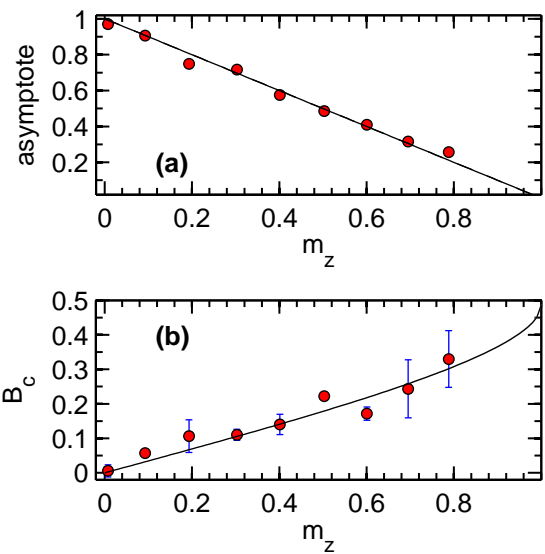}
\caption{(Color online)  {\bf a: }  Asymptotic value of $n_{0}$ for large $q$ (determined from $A_0+A_1$ in Eq.~\ref{Eq:fitfunction}). The solid line shows the value $1-m_z$ expected at zero temperature. {\bf b: } Measured critical field $B_c$ versus magnetization. The solid line shows the values expected from Eq.~(\ref{Eq:qc}) and $q_c=q_B B_c^2$, using $U_{s}/h \approx 65.6~$Hz. The gray area show the uncertainty on the theoretical value of $B_c$, dominated by the 15~\% uncertainty on the spin-dependent scattering length $a_s$. For both plots, vertical error bars show statistical uncertainties on the measured values (one standard deviation).}
\label{Fig:Bc}
\end{figure}

We measured the critical line separating the AF and BA phases using the following procedure. We bin the data according to the measured magnetization, in bins of width $0.1$ around an average magnetization from $m_z\approx 0$ to $m_z\approx 0.8$, with residual fluctuations around $\delta m_z\approx 0.02$. Each dataset with given magnetization is fitted with a function of the form 
\begin{eqnarray}
\label{Eq:fitfunction}
n_0 & =\left\{
\begin{array}{l}
A_0, q < q^\ast  \\
A_0 + A_1 \frac{q-q^\ast}{q-q^\ast+\Delta q}, q \geq q^\ast.\\
\end{array}\right.
\end{eqnarray} 
This form ensures the existence of a sharp boundary determined by $q^\ast$, a constant background value for low $q$ and a well-defined asymptotic value for large $q$, and reproduces the observed data fairly well, as shown in Fig.~\ref{Fig:dataMz05} for a specific example with $m_z\approx 0.4$. At low fields, $n_0$ is not strictly zero but takes values of a few percents, which can be explained by the presence of a small non-condensed fraction ($f'\approx 2-3~$\% per component). As such small populations are near our detection limit ($\sim3~$\% for the fractional populations, limited by the optical shot noise associated with the imaging process), we do not attempt to determine them and consider in the following that the condensate is essentially at zero temperature. At high fields, $n_0$ is very close to the expected value $1-m_z$ (see Fig.~\ref{Fig:Bc}a), again within a few percents. 

We show in Fig.~\ref{Fig:Bc}b the measured boundary between the two phases, which we find in good agreement with the prediction of Eq.~(\ref{Eq:qc}) in the whole range investigated. The comparison is made with the value $U_{s}/h \approx 65.6~$Hz, obtained from a numerical solution of the Gross-Pitaevskii equation using the scattering lengths given in \cite{knoop2011a} and the measured trapping parameters and average atom number, and thus does not require any fitting parameter. Our results are in line with previous measurements in \cite{liu2009a}, which were restricted to the range $m_{z}>0.5$ and $B> 0.2~$G and performed with much larger samples well in the Thomas-Fermi regime. Here, we are able to characterize this transition down to zero magnetization and zero applied field, in a system where spin domains (as observed in \cite{liu2009a} during the relaxation towards equilibrium) are not expected to form. 

Mean-field theory also quantitatively describes our data above the critical line. We compare the calculated $n_0$ directly to the data in Fig.~\ref{Fig:PhaseDiagram}a and b. There is no adjustable parameters in this comparison, since the parameters used in the theory are either measured or computed independently. The shape and magnitude of the  calculated phase diagrams matches well the measured one (within $10~$\% at worst), except very close to the origin $B\approx 0$ and $m_z\approx0$. In this corner of the phase diagram, we observe larger deviations from the mean field prediction and correspondingly higher fluctuations in $n_0$. We will present detailed study on these findings, which go beyond the scope of the present paper, in another publication.

In conclusion, we have explored experimentally the phase diagram of spin 1 BECs with antiferromagnetic interactions. Two phases are found, reflecting the competition between the spin-dependent interactions and the quadratic Zeeman energy. The measurements are in quantitative agreement with mean-field theory, which quantitatively predicts the phase boundary but also the observed spin populations above the transition. One expects much larger relative changes either at very low temperatures, where the interplay between spin-dependent interactions and quantum depletion are predicted to be important \cite{phuc2011a,kawaguchi2012a}, or conversely at higher temperatures where the thermodynamics should be substantially different from than of the scalar gas \cite{pasquiou2012a,zhang2004a}. Both paths provide interesting directions for future work.
 
\begin{acknowledgments}
We thank B. Laburthe-Tolra, O. Gorceix  and P. Lett for useful discussions. This work was supported by IFRAF, by Ville de Paris (Emergences project) and by DARPA (OLE project). 

\end{acknowledgments}

\newpage
\appendix
\section{Supplementary Material}

\subsection{Sample preparation}
In this section, we give a more comprehensive account of the preparation sequence used in the experiment. Evaporative cooling is done in two steps as explained in \cite{jacob2011a}, starting from a large-volume optical trap that is subsequently transferred to a smaller trap with tighter confinement (which is used for the experiments described in the main text). This sequence allows one to maintain a high collision rate throughout the whole evaporation ramp. We start from atoms loaded in the large-volume trap from a magneto-optical trap (MOT). The loading is done at a reduced trap laser power, which was found in \cite{jacob2011a} to optimize the loading. After all MOT lasers are switched off, the large-volume trap is compressed by ramping up the laser power in 2~s. This increases the collision rate in the arms of the trap, helps filling the crossing region and overall provides a better starting point for the subsequent evaporative cooling ramp. The laser cooling sequence before the compression is found to result in a mixed spin state, with spin populations in the Zeeman states $m=+1,0,-1$ in a ratio $0.7:0.2:0.1$, approximately. 

\begin{figure*}[h!!!!!!!]
 \includegraphics[width=18cm]{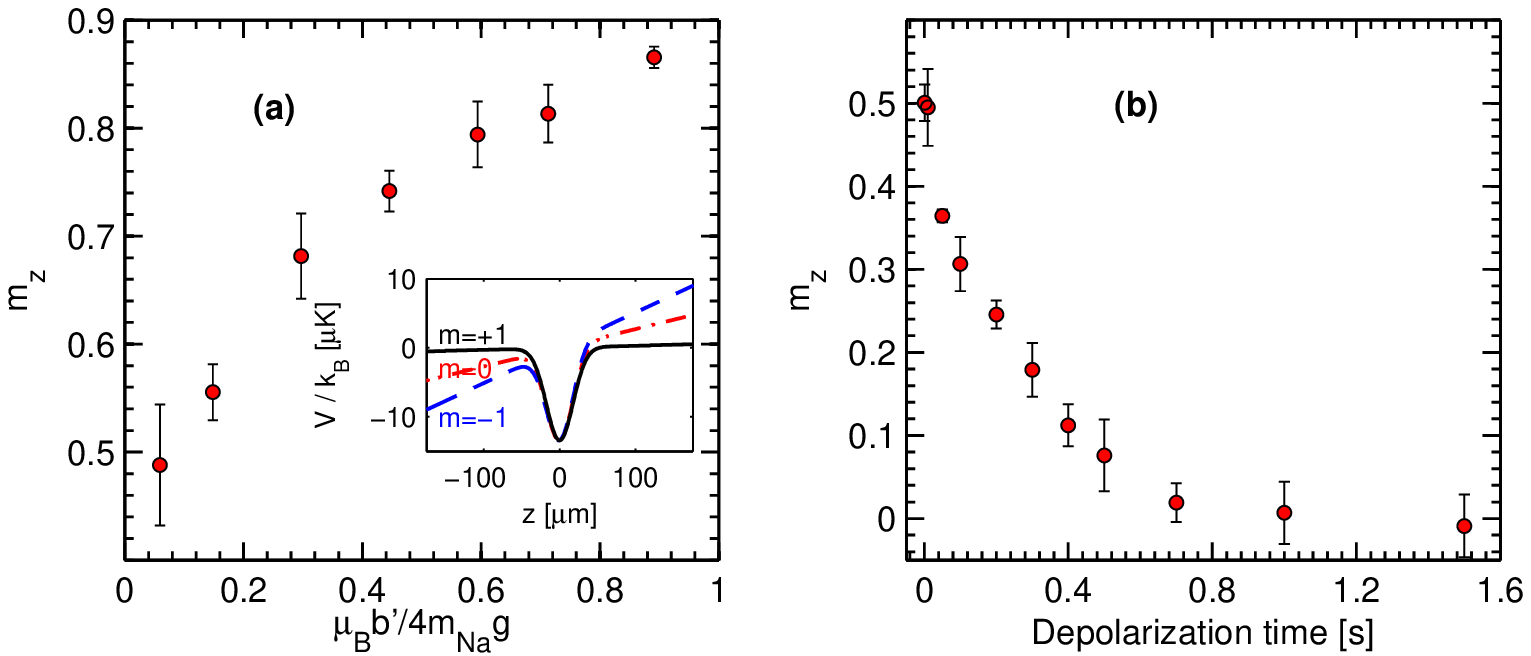}
\caption{{\bf Supplementary Material-} (Color online) {\bf (a):} Spin distillation to prepare samples with high magnetization ($m_z>0.5$). The plot shows the magnetization measured for cold clouds, as a function of the magnetic gradient $b'$ in units of $m_{Na} g/\mu_B$, with $g$ the acceleration of gravity and $\mu_B$ the Bohr magneton. The inset shows a sketch of the potential energies for each Zeeman state along the vertical axis $z$. The potential drop is exaggerated for clarity, and is smaller than depicted in the actual experiment. {\bf (b):} Depolarization to prepare samples with low magnetizations ($m_z<0.5$). The time shown corresponds to the length of a radio-frequency pulse at the Larmor frequency.}
\label{Fig:Mz_prep}
\end{figure*}

To increase the degree of spin polarization, the compression ramp is done with an additional vertical bias field ($\sim 0.5~$G) and magnetic field gradient ($20~$G/cm). As shown in \cite{couvert2009a}, this results in a spin distillation which polarizes the sample into the $m=+1$ state. The reason is that the trapping potential in the vertical direction are now slightly different for each Zeeman state, due to the potential drop caused by the gradient (see inset of Fig.~\ref{Fig:Mz_prep}). One can choose a value such that the magnetic potential almost compensates gravity for the $m=+1$ state. The $m=0$ state still feels the gravitational potential, and the $m=-1$ state then feels a potential drop twice as strong as $m=0$. As a result, the effective potential depths for $m=0,-1$ are slightly reduced compared to the $m=+1$ state, and evaporative cooling removes the former atoms preferentially. After the spin distillation is complete, we obtain a partially polarized cloud with $m_z$ ranging from $\approx 0.6$ to $\approx 0.85$ depending on the strength of the magnetic field gradient (see Fig.~\ref{Fig:Mz_prep}a). We found that keeping the gradient for longer times was no longer effective to increase the polarization further. Our interpretation is that as the cloud size becomes too small, the magnetic potential drop becomes almost unnoticeable. 

To obtain lesser degrees of polarization, we apply a horizontal bias field $\sim 0.25~$G and apply an additional radio-frequency (rf) field resonant at the Larmor frequency for a variable amount of time. As the atoms move and collide in the dipole trap, their spins quickly decohere, and produce a spin-isotropic mixture. By adjusting the strength of the rf field, we can adjust the final magnetization at will, as shown in Fig.~\ref{Fig:Mz_prep}b. The radio-frequency resonance is about $2~$kHz wide, presumably limited by inhomogeneous broadening and stray magnetic fields (which are estimated to a few mG due to environmental noise). To ensure that small frequency drifts do not perturb significantly the preparation, the frequency of the oscillating field is swept over $20~$kHz at a slow rate during the whole depolarization sequence.

After this preparation stage, we perform evaporative cooling by reducing the depth of the crossed dipole trap until a temperature $\sim 10~\mu$K is reached, at which point the cloud is transferred to the final trap with tighter focus to boost the spatial density \cite{jacob2011a}. This final trap is formed by two red-detuned laser beams, one propagating vertically with a waist ($1/e^2$ radius) of $\approx 8~\mu$m and the other propagating horizontally with a waist $\approx 11~\mu$m. At the end of the evaporation ramp, where the experiments are performed, the trap frequencies are $\{ \omega_{x,y,z} \} =2\pi\times (910,1000,425)~$Hz.

\subsection{Stern-Gerlach expansion}

\begin{figure*}[h!!!!!!!!]
\includegraphics[width=18cm]{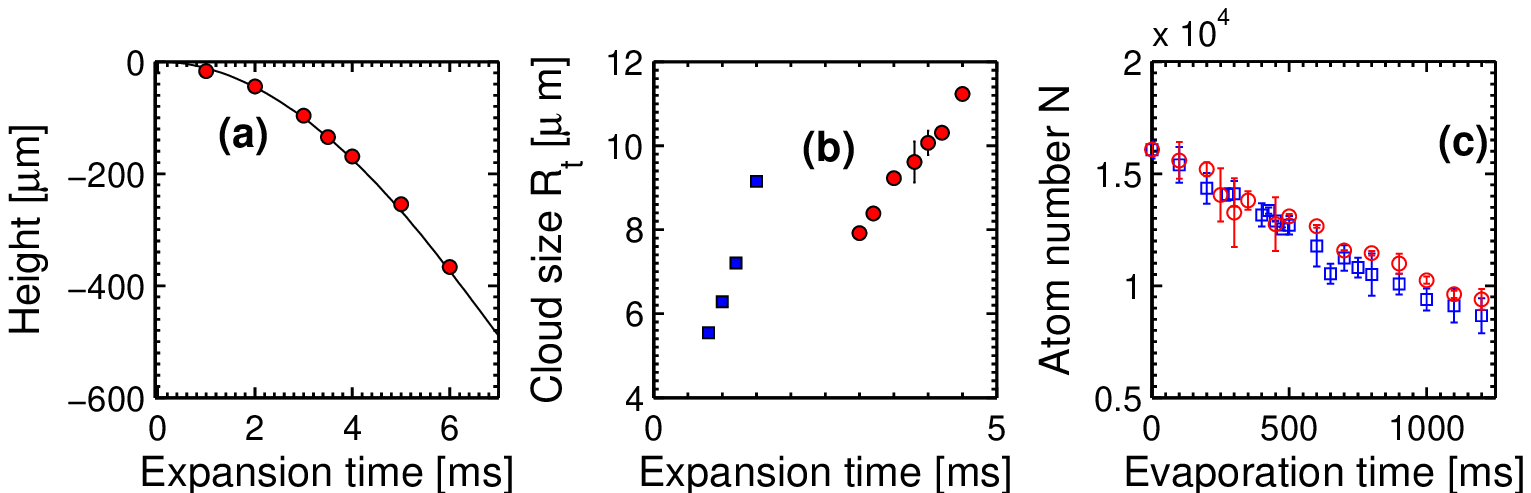} 
\caption{{\bf Supplementary Material-} (Color online) Free expansion after trap opening. {\bf (a):} Center-of-mass trajectory along the vertical direction $z$ for the $m=+1$ state; The solid line shows the calculated trajectory taking the measured magnetic gradient and gravity into account. {\bf (b):} Sizes of the expanding clouds after an expansion time for an instantaneous release (blue squares) or a smooth release (red circles). {\bf (c):} Atom number measured for instantaneous (blue squares) and smooth (red circles) releases, for various evaporation times. }
\label{Fig:ToF}
\end{figure*}

The populations of the Zeeman states are analyzed after expansion in a magnetic field gradient $b'=15~$G/cm. With an additional bias field $B_{x}\approx2~$G, this produces a force along the horizontal $x$ axis that separates the $m=\pm 1$ clouds from the $m=0$ one by a distance $d_{SG}=\mu_{B}\eta b't^2/4M_{Na}$, with  $\mu_{B}$ the Bohr magneton and $t$ the expansion time. The factor $\eta$ takes into account the temporal profile of the gradient, which rises in a few ms after the beginning of the expansion. Fig.~\ref{Fig:ToF}a shows the vertical trajectory of the atoms, and compares it with the one calculated from the measured gradient variations. The excellent agreement indicates that the gradient behavior is well understood. After a given expansion time (typically $t\approx 3.5~$ms), we take an absorption picture of the clouds, and count the relative populations. In order to obtain reliable images, the separation $d_{SG}$ must be much larger than the cloud sizes $R_{t}$ after expansion to clearly separate each Zeeman component. In our experiment, when the trap is switched off instantaneously, we typically achieve $d_{SG}/R_{t}\approx 1$ only. This is due to the tight trap frequencies, and the resulting fast expansion. The gradient strength cannot be increased further due to technical limitations, and the  expansion time is also limited by the necessity to keep a sufficiently large signal-to-noise ratio to detect atoms in each component. 

We thus resort to a slow opening of the trap, by ramping down the laser intensity to approximately $1/10$th of its initial value in $5.5~$ms before switching it off abruptly. As shown in Fig.~\ref{Fig:ToF}b, this reduces the expansion speed (of both the condensate and the thermal gas). At the same time, this leaves time for the gradient to settle to its maximum value, leading finally to $d_{SG}/R_t\approx10$ for an expansion time $t=3.5~$ms. We have checked that this procedure do not affect the measured atom number (see Fig.~\ref{Fig:ToF}c) or condensed fraction. 

\subsection{Spin interaction energy}

The spin-spin interaction energy (positive for antiferromagnetic interactions) is given in the SMA by
\begin{equation}
U_{s} = \frac{4 \pi \hbar^2 N a_{s}}{m_{\rm Na}}\int  \vert \overline{\phi}({\bf r})\vert^4~d^{3}{\bf r},
\end{equation}
 with $m_{\rm Na}$ the mass of a Sodium atom, $a_{s}\approx 0.1~$nm the spin-dependent scattering length \cite{knoop2011a} and $\overline{\phi}$ the single-mode wave function. We obtain the latter by solving numerically the Gross-Pitaevskii equation \cite{yi2002a}
 \begin{equation}
- \frac{\hbar^2}{2 m_{\rm Na} }\Delta \overline{\phi}({\bf r}) + \frac{1}{2}m_{\rm Na}\overline{\omega}^2 \overline{\phi} + N\overline{g} \vert \overline{\phi} \vert^2\overline{\phi} = \mu \overline{\phi}.
 \end{equation}
We assumed the real, slight anisotropic trap potential could be approximated by an isotropic harmonic potential, with $\overline{\omega}/2\pi\approx 0.7~$kHz the geometric average of the three trap frequencies. The spin-independent interaction strength is $\overline{g}=4\pi \hbar^2 \overline{a}/m_{\rm Na}$, with $\overline{a}\approx 2.79~$nm \cite{knoop2011a}, and in accordance with the single-mode assumption we have neglected spin-dependent interaction terms of order $\sim a_s/\overline{a}$.

\subsection{Conservation of magnetization} 

We discuss in this section the key assumption behind this work, the conservation of longitudinal magnetization. As already discussed, this is true as far as short-range interactions are concerned. However, there are other weak effects that could in principle relax the magnetization. Two main effects come to mind. First, a dipole-dipole interaction exists in principle between atoms with non-zero spin. These effects are very weak compared to short-range spin-dependent interactions. A typical relaxation rate due to dipole-dipole interactions  is less than $0.02~$Hz for a fully polarized gas for our parameters, assuming the dipolar loss constant is given by the upper bound $L_2\approx 5\cdot 10^{-16}~$at.cm$^2$/s given in \cite{goerlitz2003a}. Dipolar relaxation can therefore be neglected for the experiments reported here. Second, the cloud held in the optical trap is subject to permanent evaporative cooling, leading to a $1/e$ lifetime around $10~$s. If the potential depends on the internal state, in general magnetization is not conserved (this is the principle behind spin distillation \cite{couvert2009a}). However, even in this case the remaining trapped atoms will relax to a new equilibrium state through magnetization-conserving collisions, and thus to the equilibrium state expected with fixed magnetization. In other words, assuming the spin degrees of freedom equilibrate faster than the magnetization relaxes, the system should adiabatically follow the slow relaxation of magnetization due to evaporation. We note that if the potential is spin-independent, and the thermal gas isotropic the magnetization will not change on average, although one can expect fluctuations to increase in time. Experimentally, we found no evidence for these effects, which we believe to exist but lie beyond our sensitivity (a few percent, limited by the optical shot noise in the detection process).
\bibliography{SpinorPhaseDiagram.bib}
\bibliographystyle{apsrev}


%
%
\end{document}